\documentclass[twocolumn]{aastex62}

\usepackage{natbib}

\received{January 25, 2018}
\revised{March 19, 2018}
\accepted{April 4, 2018}
\submitjournal{ApJ}

\shorttitle{Statistical analysis of solar active regions}
\shortauthors{Rabello-Soares et al.}

\begin{document}

\title{Statistical analysis of acoustic wave power and flows around solar active regions}

\correspondingauthor{M. Cristina Rabello-Soares}
\email{cristina@fisica.ufmg.br}

\author[0000-0003-0172-3713]{M. Cristina Rabello-Soares}
\affil{Physics Department, Universidade Federal de Minas Gerais, Belo Horizonte, MG 30380, Brazil}

\author{Richard S. Bogart}
\affil{W. W. Hansen Experimental Physics Laboratory, Stanford University, Stanford, CA 94305, USA}

\author{Philip H. Scherrer}
\affil{W. W. Hansen Experimental Physics Laboratory, Stanford University, Stanford, CA 94305, USA}




\begin{abstract}

We analyze
the effect of a sunspot in its quiet surroundings 
applying a helioseismic technique on
almost three years of
Helioseismic and Magnetic Imager (HMI) observations obtained during 
solar cycle 24
to further study the sunspot structure below the solar surface.
The attenuation of acoustic waves 
with frequencies lower than 4.2 mHz depends 
more strongly
on the wave direction
at a distance of $6\degr-7\degr$ 
from the sunspot center.
The amplification of higher frequency waves is highest
6$\degr$ away from the active region and it is largely independent of the wave's direction.
We observe a mean clockwise flow around active regions, 
which
angular speed decreases exponentially with distance and has a coefficient close to $-0.7$ degree$^{-1}$.
The observed horizontal flow in the direction of the nearby active region 
agrees with a large-scale circulation around the sunspot in the shape of cylindrical shell.
The center of the shell 
seems to be centered around 7$\degr$ from the sunspot center,
where we observe an inflow close to the surface down to $\sim$2 Mm, 
followed by an outflow at deeper layers until at least 7 Mm.

\end{abstract}

%



\section{Introduction}

As acoustic waves are modified by magnetic field, helioseismology has the potential to
study the structure of sunspots below the solar surface.
There have been many 
observational and theoretical studies
that have yielded a wealth of information, but, as basic features of sunspots are still being disputed,
more work is needed \citep[see references in][]{2016ApJ...822...23Z}.	
In our previous work \citep{2016ApJ...827..140R}, 
we have analyzed the effect of a sunspot in its quiet surroundings
(i.e., neighboring regions with a small mean line-of-sight magnetic field)
several degrees away from the sunspot center over five years of solar cycle 24.
We observe that acoustic waves with frequency lower than 4.2 mHz 
are attenuated due to the presence of the nearby spot,
and waves with higher frequency are amplified.
At sunspots, this amplification happens 
for waves with frequency higher than the acoustic cutoff frequency ($\sim$5.5 mHz)
that can escape into the outer atmosphere,
which is known as the ``acoustic halo'' effect \citep{1992ApJ...394L..65B}.
Although waves propagating in the direction of the nearby active region are
more attenuated than those traveling perpendicular to it as expected, 
they have a similar amplification independent of their direction.
Here we analyze the variation of the wave power with distance to the nearby active region.

We also study the flows in the neighboring quiet regions around a sunspot and their 
variation with depth below the solar surface as observed by the acoustic waves.
Sunspots suppress the upward propagation of the heated plasma in their umbrae and 
present a strong radial surface outflows in penumbrae, 
known as the Evershed flow \citep{2009GeCAS..73Q.346E}. 
Penumbrae of evolved sunspot are usually surrounded by an additional outflow region, 
named moat flow \citep{1969SoPh....9..347S}, 
which is important in transporting flux away from the spot, and contributing to its decay 
\citep{1973SoPh...28...61H}. 
It is not clear if there is a 
connection between the Evershed and the moat flows 
\citep[e.g.][]{2014ApJ...790..135S}.  
\citet{2009SSRv..144..249G}  
observed that the moat outflow from the sunspot in NOAA 9787 
gets stronger with depth as far as 4.5 Mm deep at least;
and it is strongest about 25 Mm away from the sunspot center at all depths. 
\citet{2011JPhCS.271a2002F} 
found that the outflows extend until 10 Mm deep,
at a distance of $17-28$ Mm away from the sunspot center. 
Contrary to these works, \citet{2010ApJ...708..304Z} 
observed 
converging downward flows near the surface
at about 25 Mm away from sunspot center 
in active region NOAA 10953, 
and
outflows at deeper layers (between about 2 Mm and 8 Mm deep),
indicating the presence of a
mass circulation, 
which seems to satisfy mass conservation.
It has been argued that the near surface inflows might be important to keep sunspots stable 
\citep[and references within]{1979ApJ...230..905P,2008ApJ...684L.123H}.  
Deeper than 9 Mm or so, 
in contrast to previous findings of outflows 
\citep{2004SoPh..220..371H,2004ApJ...605..554K},	
recent works observed no large-scale flows
\citep{2010ApJ...708..304Z,2011JPhCS.271a2002F}. 
In conclusion,
it is still not quite clear what the overall flow
structures around the sunspot’s surface and interior look like.
This is also true father away from the sunspot center ($>$30 Mm).
\citet{2009SSRv..144..249G} 
observed a continuation of moat outflows as far as 100 Mm away (and 4.5 Mm deep)
while 
\citet{2009ApJ...698.1749H} 
detected an inflow from the surface down to 7 Mm deep
farther away from the sunspot center 
than the moat outflow.
\citet{2009ApJ...698.1749H} 
theorize that the 
surface cooling within the plage around the sunspot results in a downdraft, which drags fluid in at the surface,
and
the observed outflow below a depth of 10 Mm \citep[e.g.][]{2004SoPh..220..371H}   
is the return flow 
of a large-scale circulation within active regions. 
In \citet{2016ApJ...827..140R}, 
we observed inflows down to 1 Mm,
outflows below 4 Mm, and no net horizontal flow in-between,
which agree, in general, with the picture proposed by \citet{2009ApJ...698.1749H}.
Here we explore the horizontal flows of this circulation pattern with depth and
distance from the sunspot center.

In the next section, we describe the data and method used. 
Next, we present our results
first 
for
the variations in the wave amplitude and,
next for
the horizontal flows 
in the quiet regions around an active region.
Finally, we summarize our findings.

\section{Used Data and Analysis}

We use the Dopplergrams obtained 
by the Helioseismic and Magnetic Imager (HMI) on board the Solar Dynamics Observatory
\citep{2012SoPh..275..229S} 
from 2012 February to 2014 December, corresponding to Carrington rotations $2120-2157$.
The acoustic mode parameters 
are from the 
HMI Ring-Diagram processing pipeline for 5$\degr$ patches calculated every 2.5$\degr$ in latitude and longitude
and
tracked at the Carrington rotation rate for 9.6 hours 
\citep{2011JPhCS.271a2008B,2011JPhCS.271a2009B}.
The power spectrum of each patch was fitted using `rdfitc' \citep{1999ApJ...525..517B} 
and `rdfitf' \citep{2000SoPh..192..335H} methods.
In the later, the spectrum is filtered, prior to the fitting, 
to reduce any anisotropy in acoustic power
and simultaneously has its resolution decreased to increase processing speed,
as a result the number of fitted modes is smaller than the first method.
The first method measures the anisotropy in the mode amplitude:
$A = A_c \exp[ \slantfrac{A_2}{2} \cos(2\theta) + \frac{A_3}{2} \sin(2\theta) ]$
where $A_c$ is a constant isotropic term.

\begin{figure}[ht!]
\plotone{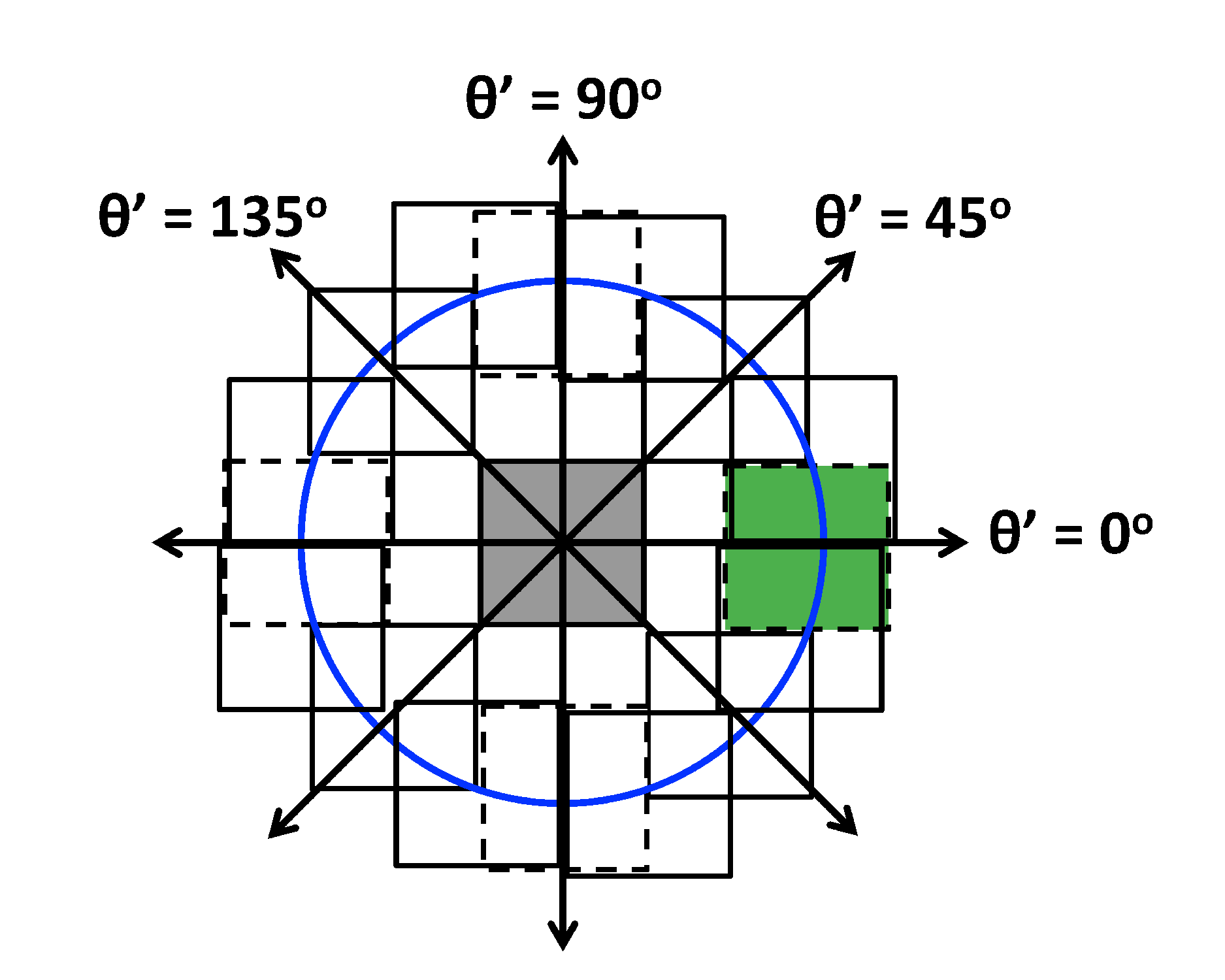} 
\caption{\label{tiles_geometry}
In the center, the grey square represents a five-degree quiet target patch.
The white squares correspond to the outermost positions that 
a neighboring active tile could be, i.e., their center are at a distance equal or smaller than 8$\degr$ (blue circle) from the center of grey square.
The white squares are slightly misplaced for a better visualization.
Assuming that the nearby active region is at the green tile (MAI $> 100$ G),
four different directions in relation to the nearby active region,
$\theta' = \theta - \theta_{ar}$,
are represented by arrows;
which will be referred to in Figure~\ref{fig2}.
The anisotropy in the mode amplitude measured by the rdfitc fitting method 
does not distinguish $\theta$ from $\theta + \pi$.
}
\end{figure}

With 
the aim to study the effect of the sunspot 
in its surroundings, we compare the acoustic wave parameters in a quiet region
with and without an active region close by.
A quiet region is defined as a region with a 
Magnetic Activity Index MAI $< 5$ G and an active region with a MAI $> 100$ G.
The MAI is calculated by the HMI processing pipeline for each tracked region using 
the equivalent 
HMI line-of-sight magnetograms
\citep{2011JPhCS.271a2008B,2011JPhCS.271a2009B}.
A five-degree tile with a MAI larger than 100 G includes NOAA active regions with a total corrected area, estimated by USAF/NOAA, as small as 40 millionths of the solar hemisphere.
A quiet region that has an active region nearby usually also has several regions with MAI $> 5$ G in its vicinity.
Thus
the distance from the quiet target tile to the nearby active region (center to center), $d_{ar}$,
is defined 
as the average of the distances to all tiles with MAI $> 5$ G in the neighborhood of the target tile weighted by their MAI
and, similarly, the angle (in relation to the solar equator), $\theta_{ar}$,
as the average of the angles to all tiles with MAI $> 5$ G in the neighborhood of the target tile weighted by their MAI
and their distance from the target.
We consider only patches with its center $8\degr$ 
(about 97 Mm)
away or less from the center of the quiet target tile
as the influence of the active region sharply falloff with distance.
Figure~\ref{tiles_geometry}
shows the neighboring tiles farthest from the quiet target tile
where an active region could be.

The relative difference of each fitted parameter ($p_j$) for a given quiet target tile $i$ is 
calculated by dividing the target tile parameter, $p_j(n,l,i)$,
by an average of the fitted parameters for 2 to 6 quiet tiles at the same disk position 
(to avoid systematic variations due to geometric foreshortening)
with only quiet tiles nearby,
and subtracting one:
$\Delta p_j / p_j ~ (n,l,i)$,
where $n,l$ are the mode radial order and degree respectively.
The parameters relative differences 
are averaged in distance, $d_{ar}$,  
mode by mode, 
over one-degree intervals,
$\langle \Delta p_j / p_j ~ (n,l,d_{ar})\rangle$,
for $d_{ar}$ = 5$\degr$, 6$\degr$, 7$\degr$, and 8$\degr$, after removing outliers greater than 3$\sigma$.
Since the analyzed regions have a five-degree diameter, a distance of 5$\degr$ means that 
the mode parameters were obtained at a quiet region 
that is next to an active region. 
The relative differences are also averaged over 
forty-five-degree intervals in $\theta_{ar}$ for each mode ($n,l$),
removing outliers greater than 3$\sigma$:
$\langle \Delta p_j / p_j ~ (n,l, d_{ar}, \theta_{ar})\rangle$.
To check the systematic errors and the level of noise introduced by our analysis,
the parameters of one of the comparison tiles (a quiet tile with only quiet tiles around it) 
for each target tile is randomly chosen and used instead of the target parameters 
and the calculations were repeated. These will be referred to as the control set.

\section{Results}

\subsection{Amplitude}

\begin{figure*}[ht!]
\plotone{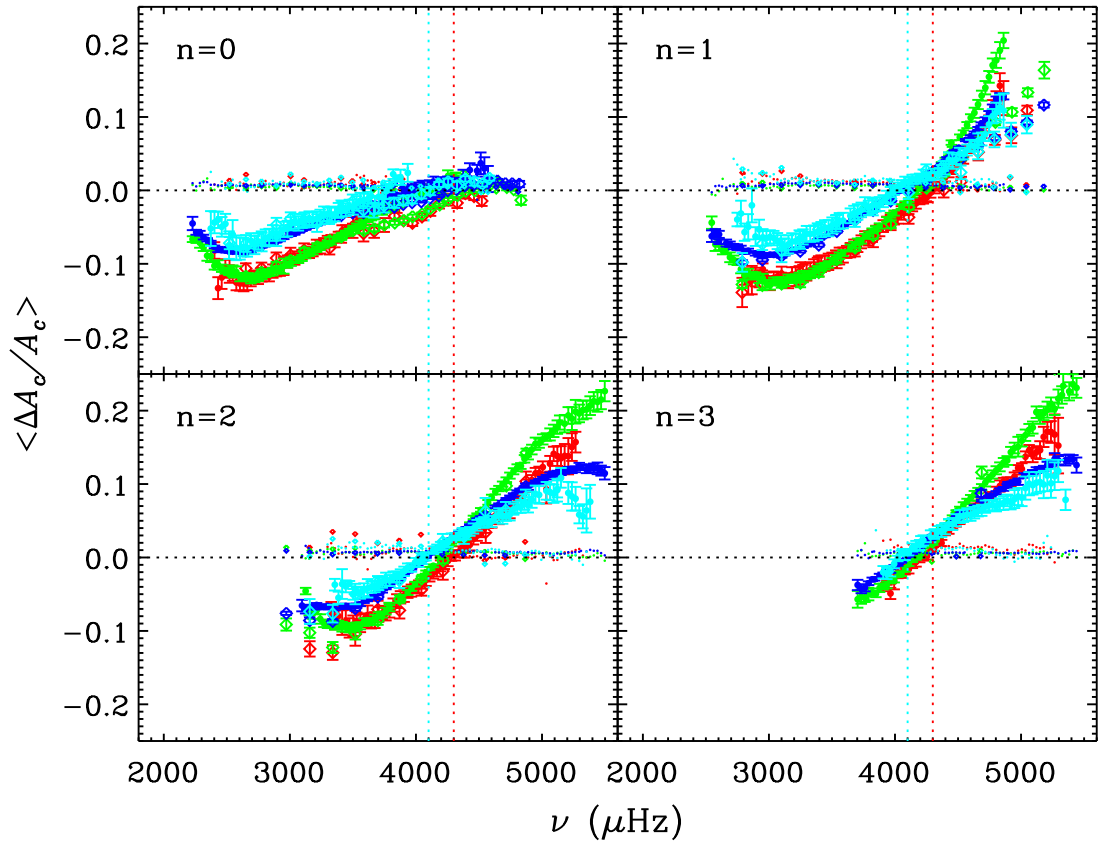}
\caption{\label{fig1}
The relative variation of the mode amplitude, $A_c$, for each mode order $n$
obtained by rdfitc (full circles) and rdfitf (diamonds).
The errors are given by the standard error of the mean.
Each color is for a given distance, $d_{ar}$, to the center of the nearby active region:
5$\degr$ (red), 6$\degr$ (green), 7$\degr$ (blue), and 8$\degr$ (cyan).
The small symbols show the values obtained for the control set and are close to zero.
}
\end{figure*}

The isotropic component of the mode amplitude, $A_c$, is presented in Figure~\ref{fig1}. 
It depends strongly on the mode frequency and slightly on mode order $n$. 
The variations observed at different distances are larger than their errors and 
the correspondent control-set variations 
(represented by tiny symbols in Figure~\ref{fig1}).
The amplitude attenuation observed closer to the active region, at $d_{ar} = 5\degr$, is not the largest as could be expected, 
but it is the same as at 6$\degr$.
There is a noticeable decrease in the attenuation between 6$\degr$ and 7$\degr$.
At 8$\degr$, it is even smaller,
which is consistent with decreasing to zero at just over 10$\degr$ away from the active region center.
The power enhancement, that happens for modes with frequencies higher than $\sim$4.2 mHz, 
is largest at 6$\degr$.
It is smaller at 7$\degr$ than at 5$\degr$ and even smaller at 8$\degr$.
For example, for modes with $n=2$ and frequency $\nu$=5 mHz, the relative amplitude difference 
is 0.12 at 5$\degr$, 0.16 at 6$\degr$, 0.11 at 7$\degr$, and 0.09 at 8$\degr$, 
which corresponds respectively to 75\%, 100\%, 69\%, and 56\% in relation to the largest variation at 6$\degr$.
%
%
The mode amplitude enhancement is known to be larger at a high inclined magnetic field 
\citep[e.g.][]{2011SoPh..268..349S,2013SoPh..287..107R}. 
%
Since the magnetic lines become more inclined as we move away from the sunspot umbra, the amplitude enhancement 
should increase with distance, but, at the same time as we move away, the magnetic field becomes weaker.
Thus 6$\degr$ away seems to be where the mode amplitude is largest when averaging over a large number of sunspots.
There is 
some indication that the change in the variation sign (from amplitude attenuation to enhancement) 
happens at lower frequencies as we move away from the nearby active region, 
from 4.3 mHz at 5$\degr$ to 4.1 mHz at 8$\degr$
(red and cyan vertical lines, respectively, in Figure~\ref{fig1}).
Again this could be due to an increase in the inclination of the magnetic field,
which reduces the acoustic cut-off frequency
\citep{2010ApJ...721L..86R}.	

%

\begin{figure*}[ht!]
\plotone{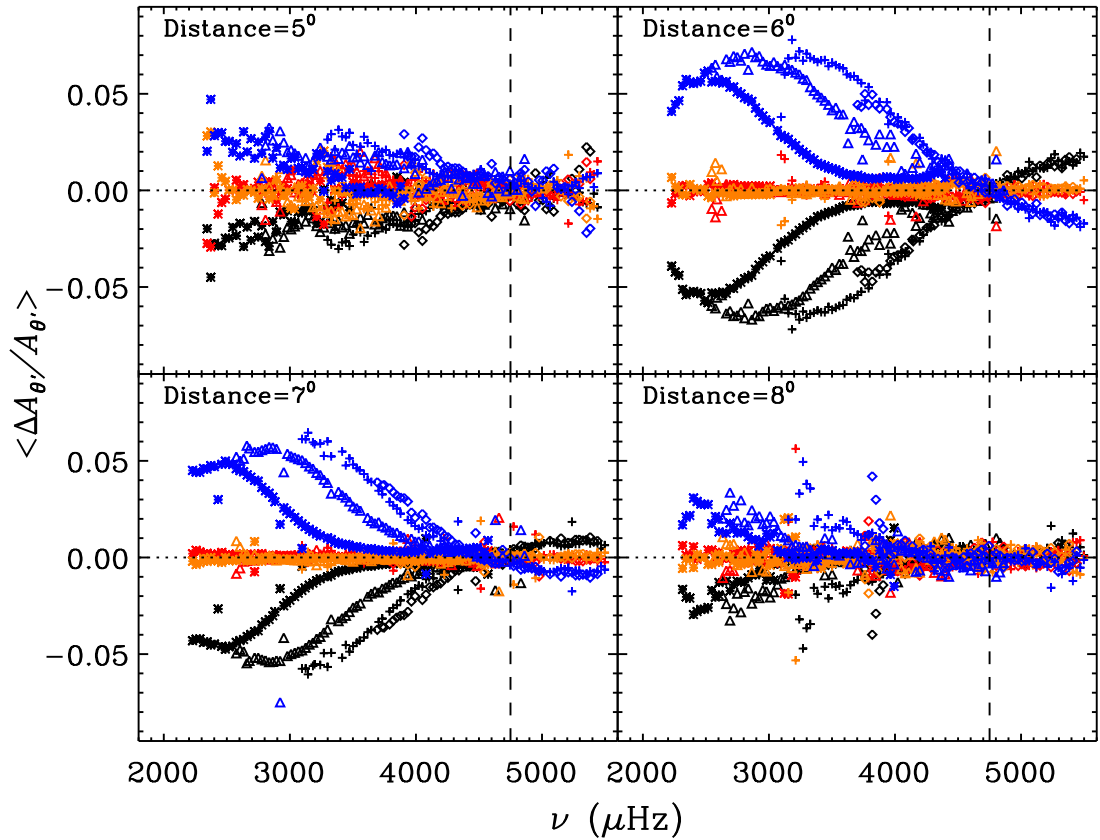}
\caption{\label{fig2}
The relative difference 
of the anisotropic component of the amplitude, $A_{\theta'}$
for four different directions:
$\theta'=0\degr$ (in the direction of the nearby active region in black),
$\theta'=45\degr$ (red), 
$\theta'=90\degr$ (perpendicular to it in blue), and
$\theta'=135\degr$ (salmon).
Besides the frequency dependence, the figure shows the variation with mode order:
$f$ (stars), $p_1$ (triangles), $p_2$ (crosses), and $p_3$ (diamonds)
modes.
}
\end{figure*}

Figure~\ref{fig2} shows
the anisotropic component of the mode amplitude variation, estimated by rdfitc method,
expressed in relation to the direction of the nearby active region, 
where $\theta' = \theta - \theta_{ar}$ (Figure~\ref{tiles_geometry}).
The anisotropic amplitude difference is largest at $6\degr$, followed by the one at 7$\degr$. 
At 6$\degr$, it varies 
12\% between waves propagating in the direction of the nearby active region 
and perpendicular to it,
which is comparable to the isotropic variation of the order of 10\% (Figure~\ref{fig1}).
Although the constant term is large close to the nearby active region, at 5$\degr$, as shown in Figure~\ref{fig1},
the anisotropic component is quite small. 
Farther away from the active region, it is even smaller,
and only noticeable for modes with frequency lower than $\sim$3.2 mHz.
There is a clear change in the sign of the variation for mode frequency around 4.75 mHz 
(vertical line in Figure~\ref{fig2}).
Low frequency modes that travel
in the direction of nearby active region are attenuated
in relation to modes propagating perpendicular
to it,
while high frequency ones are enhanced
\citep[see also Figure~2 in][]{2016ApJ...827..140R}. 
The maximum attenuation is larger than the maximum enhancement.
At 6$\degr$ and 7$\degr$, it is by a factor of approximately five.
The sign change happens at lower frequencies as the distance increases.
At 6$\degr$, it is sharp and happens at 4.75 mHz (Figure~\ref{fig2} top right panel).
At 7$\degr$, the change happens over a range in frequency, where modes with frequency 
between 4.4 and 4.7 mHz 
independent of its direction.
At 8$\degr$, modes with frequency lower than $\sim$3.2 mHz are attenuated.
At 5$\degr$, modes with frequency lower than $\sim$4.8 mHz are attenuated,
although it is more evident for frequencies lower than $\sim$4.1 mHz.

\begin{figure*}[ht!]
\plotone{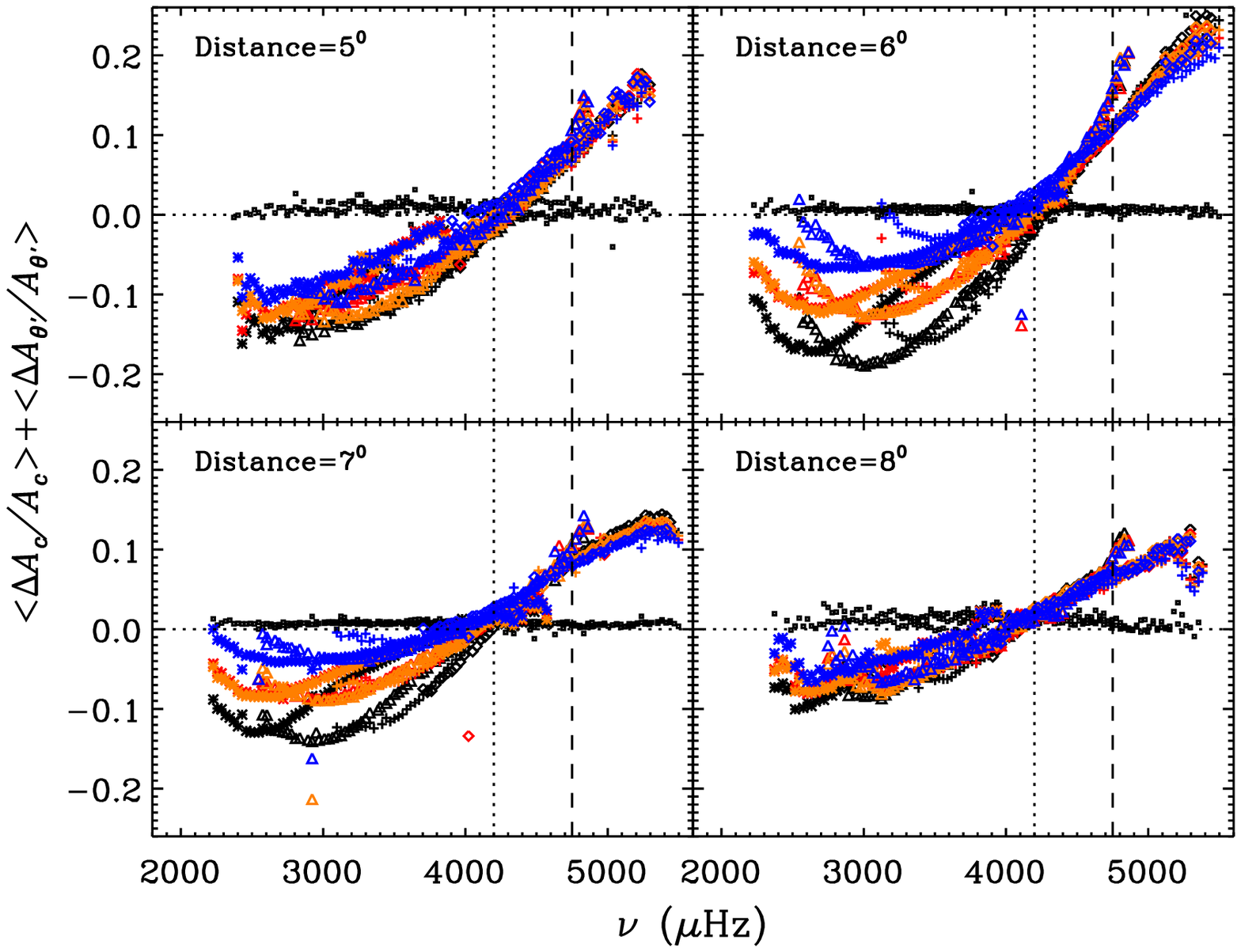}
\caption{\label{fig3}
The relative difference of the maximum power in the ring, $A_c A_{\theta'}$ 
at four different distances from the nearby active region for
$f$ (stars), $p_1$ (triangles), $p_2$ (crosses), and $p_3$ (diamonds)
modes.
The colors show different directions:
$\theta'=0\degr$ (in the direction of the nearby active region in black),
$\theta'=45\degr$ (red), 
$\theta'=90\degr$ (perpendicular to it in blue), and
$\theta'=135\degr$ (salmon).
The small black squares close to zero are the results for the control set.
}
\end{figure*}

Although modes with frequency lower than $\sim$4.75 mHz propagating in the direction of nearby active region are attenuated in relation to modes propagating
perpendicular
to it,
the ones with frequency higher than 4.2 mHz are, in fact, amplified when taking into account the isotropic term.
Putting both components of the amplitude together in
Figure~\ref{fig3}, it clearly shows that the anisotropic variation is important at 6$\degr$ and 7$\degr$ away from the nearby active region.
On the other hand, at our closest and our farthest locations, the waves are affected 
approximately in the same way
by the nearby active region independent of their direction.
At 8$\degr$, the amplitude variation is smaller than at 5$\degr$.  
The attenuation at 5$\degr$ for waves propagating in all directions is very similar to 
the one for waves propagating diagonally (at 45$\degr$ or 135$\degr$) at 6$\degr$.
In fact, 
we observe the largest attenuation for modes
propagating in the direction of the nearby active region at a 6$\degr$ distance 
(and not closer to the active region at 5$\degr$) with frequency around 3 mHz and order $n$=1. 
The attenuation at 7$\degr$ also strongly depends on the wave direction 
and it is about 50\% smaller than at 6$\degr$.
The amplification is also the highest at 6$\degr$ and increases with mode frequency.
It has very little dependence with the propagation direction at all analyzed distances.
The relative amplitude variation observed for different distances are larger than the variations observed by the control set (tiny symbols in Figure~\ref{fig3}).

\citet{2011ApJ...740...56Z}	
observed that the amplitude of the wave modified by the active region 
decreases with mode order $n$, which agrees, in general, with our results.
This is more noticeable in our results 
at 6$\degr$ and 7$\degr$ for waves propagating in the direction of the nearby 
active region where, for a given frequency, in most cases,
the amplitude is more attenuated as $n$ increases (Figure~\ref{fig3} symbols in black).
However, we observe an inversion in this behavior in some instances, for example:
for 2.6 mHz the $f$ modes are more attenuated than $p_1$ modes, and
for 3.2 mHz the $p_1$ modes are more attenuated than $p_2$ modes.

\subsection{Flow}

\begin{figure*}[ht!]
\plotone{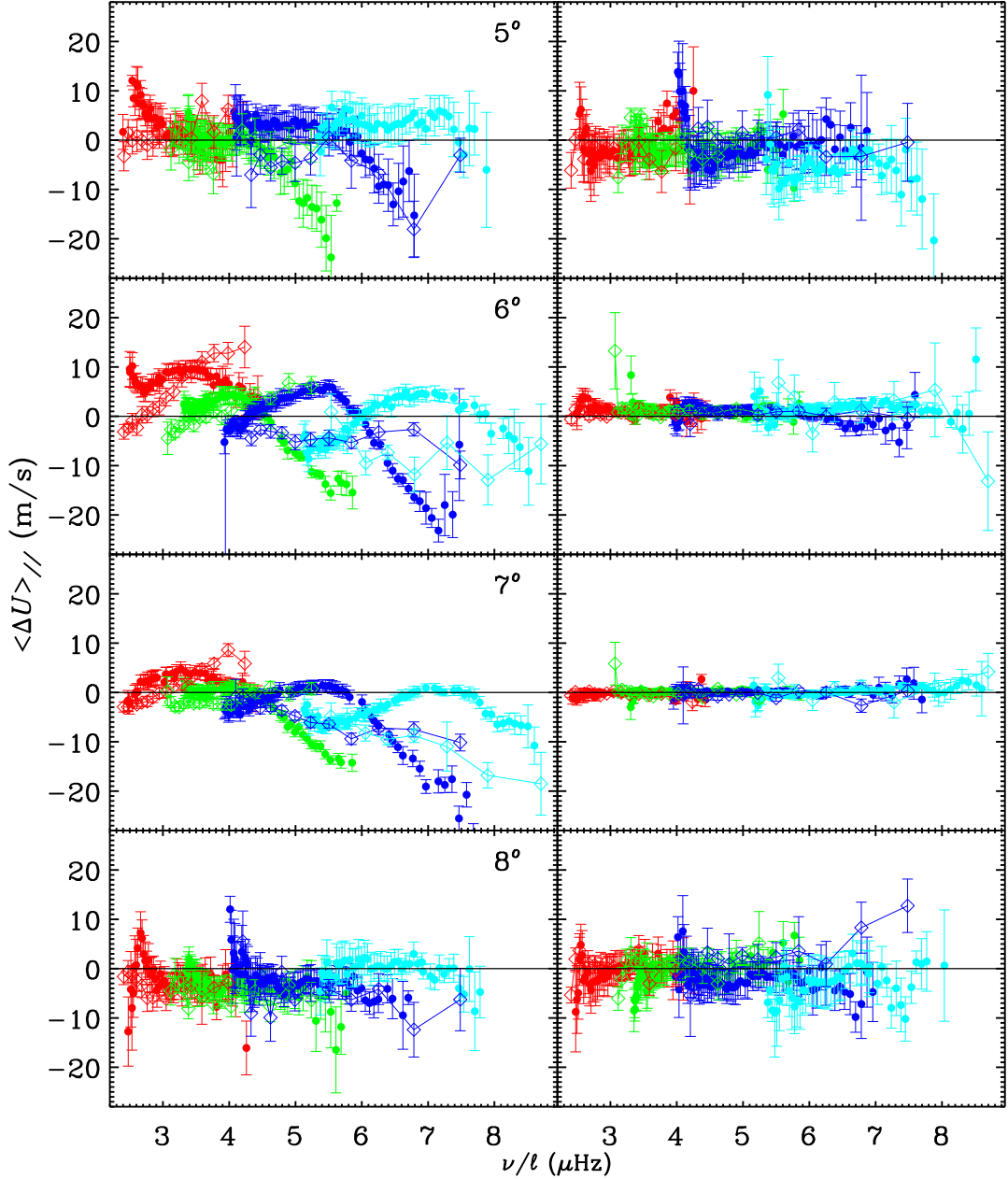}
\caption{\label{fig4}
Flow variation parallel to (i.e., in the direction of)
the nearby active region (left)
and the variation given by the control set (right)
obtained using rdfitc (circles) and rdfitf (diamonds)
for $f$ (red), $p_1$ (green), $p_2$ (blue), $p_3$ (cyan) modes.	
For better visualization, the modes obtained using rdfitf are connected by a line.
The error bars represent the standard error of the mean.
Negative flows indicate
flows moving away from the nearby active region (outflows) and positive flows moving toward the nearby active region (inflows). 
}
\end{figure*}

The flow obtained using rdfitc and rdfitf is changed from a
solar longitude-latitude reference system 
to a parallel-perpendicular to the nearby active region system.
The left panels in Figure~\ref{fig4} show the flow variation parallel to the nearby active region 
as a function of the mode lower turning point,
given by its frequency divided by its angular degree $l$.
Although similar, there are differences between the fitted flows obtained by rdfitc and rdfitf.
The panels on the right in Figure~\ref{fig4} are obtained using the control set.
The estimated systematic flow variations are small only at 6$\degr$ and 7$\degr$.
For the other two distances, they are of the same order of magnitude as the observed flow
indicating that they have large systematic errors.

\begin{figure}[ht!]
\plotone{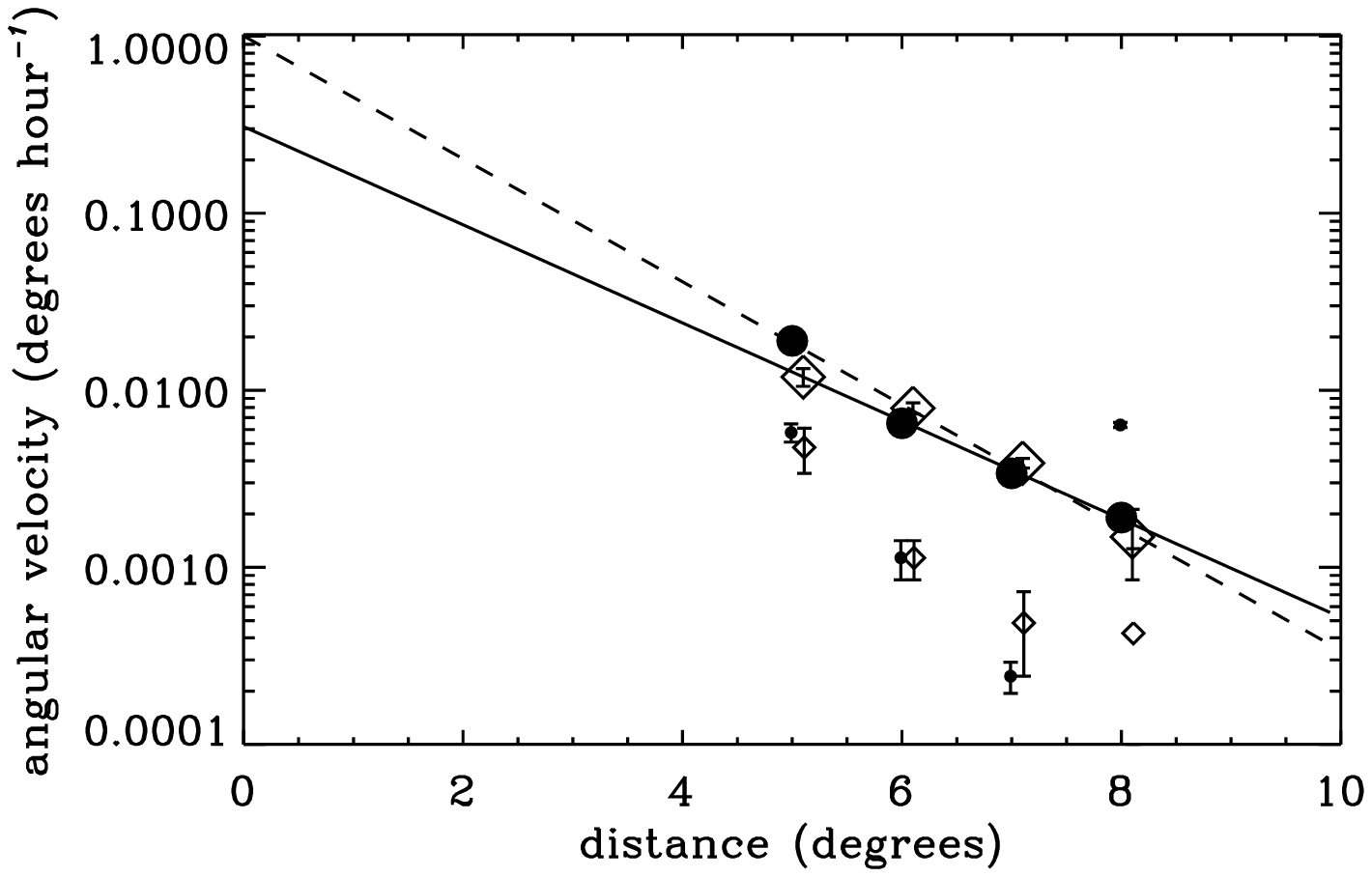}
\caption{\label{angular_speed}
Mean angular speed for rdfitc (full circles) and rdfitf (diamonds) from the flows in Table~\ref{table1}. The small symbols are for the control set.
The symbols have been slightly displaced horizontally 
to improve visibility.
The straight lines are exponential fits to rdfitc (full line) and rdfitf (dashed line) angular velocities. 
Extrapolation of the rdfitc fit (full line) roughly agrees with the average of 
angular speed obtained by \citet{2016ApJ...826....6Z} 
at the sunspot umbra 
for all rotating sunspot during almost 40\% of our data set.
}
\end{figure}

We also observe a small but significant flow perpendicular to the nearby active region.
It is very similar for all modes.	
Table~\ref{table1} shows the weighted mean over all modes for each distance and fitting method.
It can also been seen in our previous work in the right panel 
of Figure~8 \citep[in][]{2016ApJ...827..140R}. 
It is consistent with a clockwise flow around the active region and 
it decreases as the distance to the active region increases.
These results are larger than the systematic effects obtained with the control set
as shown in Table~\ref{table1}.
Some sunspots are observed to rotate around their umbral centers.
Recently,
\citet{2016ApJ...826....6Z}	
found that sunspots
tend to rotate counterclockwise 
in the southern hemisphere,
and have the opposite behavior in the northern hemisphere,
using HMI observations from 2014 January to 2015 February,
which overlaps in part with our data set.
However, using MDI data during the year of 2003, they observed the opposite behavior.
According to the 
Sunspot Index and Long-term Solar Observations graphics of the
Royal Observatory of Belgium\footnote{http://www.sidc.be/silso/monthlyhemisphericplot}
during our analyzed time period, there was excess in the sunspot number in the south hemisphere in relation to the north,
suggesting a net counterclockwise rotation.
However, averaging the estimated angular speed for all rotating sunspots obtained by 
\citet{2016ApJ...826....6Z} during 2014 (in their Table 1), 
we obtain a net clockwise angular speed of 0$\degr$.25 hr$^{-1}$ at sunspot umbra.
%
%
Our results show a decrease of angular speed with distance ($d_{ar}$) 
and fitting an exponential function: $\omega_0 \exp(c \,\, d_{ar})$,
we obtain the exponential coefficient
$c = (-0.63809 \pm 0.00003)$ degree$^{-1}$ for rdfitc (full line) and 
$c = (-0.79915 \pm 0.00007)$ degree$^{-1}$ for rdfitf (dashed line 
in Figure~\ref{angular_speed}).
At the sunspot umbra ($d_{ar} \approx 0$), 
the angular speed is:
$\omega_0 = (0\degr.30825 \pm 0\degr.00007$) hr$^{-1}$ for rdfitc and
$\omega_0 = (1\degr.0029 \pm 0\degr.0005$) hr$^{-1}$ for rdfitf,
which are of the same order as 
\citeauthor{2016ApJ...826....6Z} 
results.
%
Although 
\citet{2003ApJ...591..446Z} found a 	
strong vortex in and around a sunspot in the active region NOAA 9114
extending from the surface to 5 Mm deep
and a vortex in the opposite direction in deeper layers ($9-12$ Mm),
in our analysis, we do not observe as deep as will be discussed next,
which might explain the lack of depth dependence in our results.

\begin{deluxetable*}{cCCCC}[b!]
\tablecaption{
The weighted mean of the flow (in m~s$^{-1}$) moving perpendicular 
to the nearby active region at each distance.
Positive flow is in the clockwise direction.
\label{table1}
}
\tablecolumns{5}
\tablenum{1}
\tablewidth{0pt}
\tablehead{
\colhead{Method} &
\colhead{$5\degr$} & \colhead{$6\degr$} &
\colhead{$7\degr$} & \colhead{$8\degr$}
}
\startdata
rdfitc & $5.6\pm0.2$ & $2.3\pm0.1$ & $1.4\pm0.1$ & $0.9\pm0.1$\\
control & $1.7\pm0.2$ & $0.4\pm0.1$ & $0.10\pm0.02$ & $3.0\pm0.1$\\
\hline
rdfitf & $3.5\pm0.4$ & $2.8\pm0.2$ & $1.6\pm0.1$ & $0.7\pm0.3$\\
control & $1.4\pm0.4$ & $0.4\pm0.1$ & $0.2\pm0.1$ & $0.2\pm0.4$\\
\enddata
\end{deluxetable*}

\begin{figure*}[ht!]
\plotone{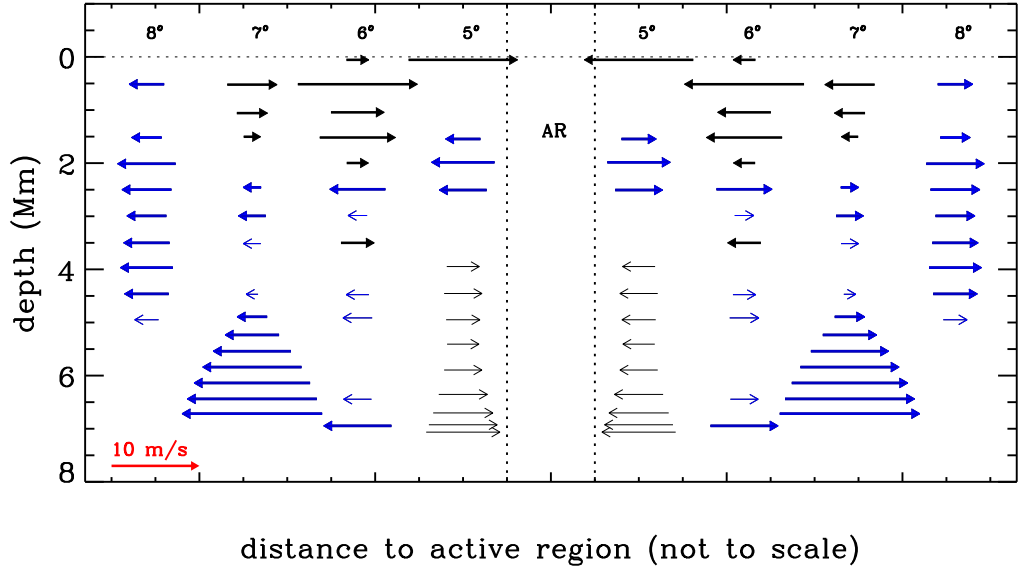}
\caption{\label{flow_inverted}
Inferred flow variation parallel to (i.e., in the direction of)
the nearby active region obtained for rdfitc.
Thick arrows are for flows larger than 2.5 $\sigma$ and thin arrows are for $1.5-2.5 \sigma$.
The size of the red arrow corresponds to a flow of 10 m~s$^{-1}$.
For better visualization, outflows are in blue and inflows in black,
plus
the results were duplicated in the left and right side of the active region.
}
\end{figure*}

\begin{figure}[ht!]
\plotone{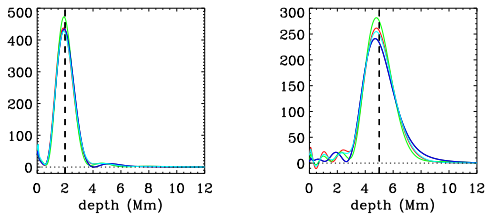}
\caption{\label{avgker}
Examples of averaging kernels at selected depths: 2 Mm (left panel) and 5 Mm (right panel).
Each color is for a given distance, $d_{ar}$, to the center of the nearby active region:
5$\degr$ (red), 6$\degr$ (green), 7$\degr$ (blue), and 8$\degr$ (cyan).
They are defined so that its integral over the solar interior is made equal to one.
}
\end{figure}

Figure~\ref{flow_inverted} presents the inferred flow parallel to the nearby active region
at different depths below the solar surface,
obtained using the optimally localized average 
\citep[OLA;][]{1968GeoJ...16..169B} technique
implemented in the HMI Ring-Diagram analysis pipeline (module rdvinv)
- for detailed information see references in \cite{2011JPhCS.271a2008B}.
%
%
%
Only the results from rdfitc were used due to its larger number of fitted modes
(214, 231, 220, 216 modes for distances 5$\degr$, 6$\degr$, 7$\degr$ and 8$\degr$ respectively),
as shown in Figure~\ref{fig4} (left panels).
The best trade-off parameter for these data set is $\mu$ = 0.00020 
\citep[e.g. Figure 9 in][]{2016ApJ...827..140R}.
The averaging kernel for two chosen depths are shown in Figure~\ref{avgker}.
%

%
%
Only inferred flows larger than 1.5 times its error were plotted.
Thick arrows are for inferred flows larger than 2.5 $\sigma$.
The solution error is, on average, $\sim$1 m~s$^{-1}$ close to the solar surface and increases
to $\sim$2 m~s$^{-1}$ at 7 Mm deep.
The depth is given by the second quartile point of the averaging kernel.
The solution is better localized close to the surface
and becomes worse with increasing depth:
$\Delta_{qu}$ = 0.6 Mm at 0.5 Mm and $\Delta_{qu}$ = 2 Mm at 7 Mm,
where $\Delta_{qu}$ is the difference between the first and third quartile points of the averaging kernel.

At 7$\degr$, for which the control flows are small (Figure~\ref{fig4} right panels),
we observe an inflow close to the surface, until $\sim$2 Mm deep, 
followed by an outflow at deeper layers.
The outflow increases with depth, reaching 15 m~s$^{-1}$ at 7 Mm.
This is also observed generally at 6$\degr$, except for some indication of a brief change of direction
around 3.5 Mm deep.
Our measurements agree with 
a large-scale circular flow 
in the shape of a ring (or a cylindrical shell) around the sunspot
as proposed by \citet{2009ApJ...698.1749H} and similar to the one proposed by \citet{2010ApJ...708..304Z},
supposing 
a downflow in the ring inner part and a upflow in the outer part
connecting the near-surface inflow and the deep outflow.
In our results, this circular flow extends as far as 7$\degr$ away from the active region center.
Although the outflows at 8$\degr$ are larger than 2.5 $\sigma$, their control set present similar outflows 
shown in Figure~\ref{fig4} as negative flows.
At 5$\degr$, the outflow is observed closer to the surface, around 2 Mm deep, 
but care should be taken in analysing these results as the control flow,
while not having a $\nu/l$ dependence, are quite large.
Our results differ from
\citet{2009SSRv..144..249G}
as
they  observed outflows in the upper 4 Mm 
until as far as 120 Mm from the center of one sunspot (NOAA 9787).
Also,
averaging their Figure~11 over a 5$\degr$ length 
to compare with our results, 
their estimated flows are much larger than ours:
at 2.5 Mm depth, their values at 5$\degr$ and 7$\degr$ are, respectively,
about 70 m~s$^{-1}$ and 20 m~s$^{-1}$ to be compared with ours: $-5$ m~s$^{-1}$ and $-2$ m~s$^{-1}$. 
\citet{2009SSRv..144..249G} results
are also in disagreement with \citet{2010ApJ...708..304Z}, 
even though both used the same helioseismic method.

\section{Summary}

We calculted the mean difference between the acoustic wave amplitude and flows observed 
in quiet regions next to an active region 
and the ones surrounded only by quiet regions 
over 38 Carrington rotations applying ring-analysis to 
5$\degr$ regions in HMI Dopplergram images. 
We analyzed the variations with distance and propagation direction from the nearby active region 
as far as 8$\degr$ away from the active region center.

{\it Amplitude.}
The mode amplitude is attenuated by the presence of the nearby active region for modes with frequency lower than about 4.2 mHz and it is amplified for higher frequencies. 
The amplitude difference depends on the direction of the wave in relation to the nearby active region.
This anisotropic variation is more important at 6$\degr$ and 7$\degr$ 
than at 5$\degr$ or 8$\degr$ away from the nearby active region.
Although the maximum enhancement observed is as large as the maximum attenuation,
it is largely independent of the wave direction.
The largest amplification happens 6$\degr$ away from the active region. 
On the other hand,
waves that travel perpendicular to the nearby active are attenuated nearly 40\% less than the ones
propagating in its direction.
The largest attenuation happens for $p_1$ modes
with frequency close to 3 mHz at 6$\degr$ away from the nearby active region
and traveling in its direction.
The anisotropic component of the amplitude changes from attenuation to enhancement at a higher frequency than the isotropic term, around 4.8 mHz. Thus, although it is a small effect, waves with frequency between $4.2-4.8$ mHz moving perpendicular to the nearby active region are more amplified than waves traveling in its direction.

{\it Flow.}
We observe a mean clockwise flow around active regions which
decreases with distance 
from close to 5 m~s$^{-1}$ to less than 1 m~s$^{-1}$.
We fitted an exponential function to the angular speed with a coefficiente close to $-0.7$ degree$^{-1}$. 
Extrapolating our fit, we get a angular speed at sunspot umbra similar to an average 
of all sunspots observed to rotate around their umbral centers
obtained by \citet{2016ApJ...826....6Z} 
during almost 40\% of our data set.
We also analyzed the variation of the flow in the direction of the nearby active region 
with distance
from near the surface to 7 Mm deep.
Our results agree with 
a large-scale circular flow 
around the sunspot
in the shape of a cylindrical shell around the sunspot
as proposed by \citet{2009ApJ...698.1749H} 
supposing 
a downflow close to the sunspot and a upflow farther away
connecting the near-surface inflow and the deep outflow.
The center of the shell of this circular flow 
seems to be centered around 7$\degr$ from the sunspot center,
where
we observe an inflow close to the surface down to $\sim$2 Mm, 
followed by an outflow at deeper layers until at least 7 Mm.


\acknowledgements

This research was supported in part by
Minas Gerais State Agency for Research and Development (FAPEMIG) and
the Brazilian National Council for Scientific and Technological Development (CNPq).

%

\vspace{5mm}
\facility{SDO(HMI)}

\bibliography{rabellosoares_bib}




\end{document}